\documentclass[12pt]{iopart}
\usepackage{iopams}
\usepackage{amsfonts}
\usepackage{amssymb}
\usepackage{setstack}

\eqnobysec

\begin{document}
%
%

\renewcommand{\a}{{\boldsymbol a}}
\renewcommand{\b}{{\boldsymbol b}}
\renewcommand{\c}{{\boldsymbol c}}
\renewcommand{\i}{{\boldsymbol i}}

\newcommand{\0}{{\boldsymbol 0}}
\newcommand{\1}{{\boldsymbol 1}}
\newcommand{\2}{{\boldsymbol 2}}
\newcommand{\3}{{\boldsymbol 3}}

\newcommand{\bj}{{\boldsymbol j}}

\newcommand{\ba}{{\boldsymbol{\alpha}}}
\newcommand{\bb}{{\boldsymbol {\beta}}}

\newcommand{\A}{{\boldsymbol A}}
\newcommand{\B}{{\boldsymbol B}}
\newcommand{\C}{{\boldsymbol C}}
\newcommand{\D}{{\boldsymbol D}}
\newcommand{\E}{{\boldsymbol E}}
\newcommand{\F}{{\boldsymbol F}}


\newcommand{\tlam}{\tilde{\lambda}}
\newcommand{\tP}{\tilde{P}}
\newcommand{\ty}{\tilde{y}}
\newcommand{\tx}{\tilde{x}}
\newcommand{\tf}{\tilde{f}}
\newcommand{\tg}{\tilde{g}}


\newcommand{\Si}[3]{\tilde{\Sigma}_{#1}^{\,\,\,#2#3}}
\newcommand{\ud}{\,\mathrm{d}}
\newcommand{\half}{\textstyle{\frac{1}{2}}}
\newcommand{\m}[1]{\mathcal{#1}}
\newcommand{\var}[2]{\frac{\delta{#1}}{\delta{#2}}}
\newcommand{\ap}{\approx}
\newcommand{\pa}{{\phantom{a}}}
\newcommand{\pd}[1]{
        {\frac{\partial}{\partial{#1}}}}


\title[2+2 Ashtekar variables II]{Hamiltonian analysis of the double null 2+2 decomposition of General Relativity expressed in terms of self-dual bivectors}
%
%
\author{R A d'Inverno, P Lambert   and J A Vickers}
\address{School of Mathematics, University of Southampton, Southampton, SO17~1BJ, UK}
\ead{J.A.Vickers@maths.soton.ac.uk}
\begin{abstract}
In this paper we obtain a 2+2 double null Hamiltonian
description of General Relativity using only the (complex) $SO(3)$ connection
and the components of the complex densitised self-dual bivectors $\Sigma_\A$. We
carry out the general canonical analysis of this system and
obtain the first class constraint algebra entirely in terms of the
self-dual variables. The first class algebra forms a Lie algebra and all 
the first class constraints have a simple geometrical interpretation. 
\end{abstract}
\submitted{\CQG}
\pacs{0420Fy}
\maketitle
%
\section{Introduction}
In two earlier papers \cite{rd&jv:95} \cite{pap1} 
(which we will refer to as papers I and II respectively) we
gave a 2+2 Hamiltonian description of General Relativity using
Ashtekar type variables based on a self-dual action
\cite{tj&ls:88}, \cite{samuel}. 
The variables used in paper II were based on previous
work by the authors \cite{rd&jv:95} and also on the approach taken by
Goldberg et al \cite{jg&dr:92} and used a mixture of frame variables and
the components of the self-dual bivector. This had the advantage that
the variables had a clear geometric interpretation but suffered from
the disadvantage that not all the variables were manifestly
self-dual. In this paper we will describe an alternative approach in
which the variables are the components of the densitised self-dual
bivectors $\Sigma_\A$ together with the components of the complex
$SO(3)$ connection 1-forms $\Gamma^\A$. As in paper II we use 
the original Ashtekar formulation in which the manifold is real but 
we allow complex solutions of the field equations. However once the
canonical analysis has been carried out reality conditions are imposed to
limit the solutions to  real solutions of Einstein's equations.

In the first section we use the self-dual action introduced in paper I
to obtain a Lagrangian description expressed in terms of these
variables. However if the the bivector variables are obtained from a
null frame the components are not independent and one needs to impose
appropriate constraints in order to obtain General Relativity. These
are found and introduced into the Lagrangian using Lagrange
multipliers. This immediately gives rise to the primary Hamiltonian and
in the next section we look at the Hamiltonian description. 

In section 4 we show that both the structure equations for the connection
and all the Einstein equations are given by the constraints and equations
of motion. We then go on in section 5 to calculate the first class
constraints and obtain the first class algebra. This is shown to be a Lie
algebra generated by the momentum constraint and the modified Gauss
constraint. We end by giving a geometrical interpretation to the
constraints and discuss the next steps in the canonical quantisation
process.

\section{2+2 Connection Variables}
As in paper I we we start with the first order self-dual action
given by Jacobson and Smolin \cite{tj&ls:88}
\begin{equation}
    I=\int R^{\A}\wedge S^{\B}g_{\A\B},
\end{equation}
where $S^{\A}$ are the complex self-dual 2-forms, $R^{\A}$ is the
curvature 2-form of the complex $SO(3)$ connection $\Gamma^{\A}$ and
$g_{\A\B}$ is the $SO(3)$ invariant metric.
If we introduce a densitised version of $S^{\A}$ by
\begin{equation}
    \Si \A\alpha\beta=\half\epsilon^{\alpha\beta\gamma\delta}
        S^\B_{\pa \gamma\delta}g_{\A\B},
\end{equation}
this  leads to the Lagrangian density
\begin{equation}
    \m{L}=R^\A_{\pa\alpha\beta}\Si\A\alpha\beta.
\end{equation}
By making a $2+2$ decomposition of the connection according to
\begin{eqnarray}
    \Gamma^\A=\Gamma^\A_{\pa \mu}dx^{\mu}
        =A^\A_{\pa i}dx^i+B^\A_{\pa a}dx^a.
\label{2.2conn}\end{eqnarray}
and using the $SO(3)$ covariant derivative $D$ this may be written as
\begin{equation}
\fl  \m{L}   =\Si \A0i A^\A_{i,0}+\Si \A01B^\A_{1,0} +B^\A_0D_1\Si \A01
        +B^\A_0D_i\Si \A0i
        +R^\A_{\pa 1i}\Si \A1i +R^\A_{\pa 23}\Si \A23 . \label{eq:Ld}
\end{equation}

In paper I we considered the phase space to be given by the twenty three
variables $\Si\A01$, $\Si \A ai$, $\mu^{\a}_{\pa b}$, $ s^{i}_{\pa a}$
together with the connection terms $A^\A_i$ and $B^\A_a$. These twenty
three variables were required to satisfy thirteen constraints which left
ten degrees of freedom: two spin and boost freedoms (self-dual and anti
self-dual) and eight degrees of freedom for the double null metric (the 10
degrees of freedom for a general metric are reduced to 8 because of the two
slicing conditions). However, as explained in the introduction, in this
paper we will work directly with the densitised bivectors and use all the
components of $\Si \A\alpha\beta$ rather than use a mixture of the sigma
variables and frame components as we did in paper II. There are a total of
18 independent components and these have to satisfy nine constraints. This
means that we have only nine degrees of freedom; one less than in paper
II. The reason for this is that the variables we are now using are all
manifestly self-dual so that we need only consider the effect of the
self-dual spin and boost transformations on the sigma variables. This is
different from paper II where the frame variables $\mu^{\a}_{\pa b}$ were
not invariant under anti self-dual transformations and this therefore had
to be considered as a gauge freedom.  However in both cases the full
Hamiltonian analysis which involves splitting the constraints into first
and second class constraints gives the same number of dynamical degrees of
freedom.

We now give the nine constraints. The first
two constraints are found by expressing the results
$S^{\1}\wedge S^{\2}=0$ and $S^{\1}\wedge S^{\3}=0$ in terms of $\Si
\A\alpha\beta$. This gives  
\numparts \label{eq:c1&2} \begin{eqnarray}
\epsilon_{\alpha\beta\gamma\delta} 
\Si \1\alpha\beta \Si \2\gamma\delta&=0\label{eq:c1},\\ 
\epsilon_{\alpha\beta\gamma\delta}
\Si \1\alpha\beta \Si \3\gamma\delta&=0\label{eq:c2}.
\end{eqnarray} \endnumparts
The next set of constraints are obtained by considering the
expressions for $\Si \A\alpha\beta$ in terms of the frame
variables. This gives the following four constraints:
\numparts\label{eq:c3-6} 
\begin{eqnarray} 
\epsilon_{ij}\Si \20i\Si \21j=0, \label{eq:c3}\\ 
\epsilon_{ij}\Si \30i\Si \31j=0, \label{eq:c4}\\
\Si\201=0,\\ 
\Si\301=0.
\end{eqnarray} 
\endnumparts
Two further constraints are found by  expressing 
$\mu_{\pa 0}^{\mathbf{1}}\mu_{\pa 0}^{\mathbf{0}}$ and
$\mu_{\pa 1}^{\mathbf{0}}\mu_{\pa 1}^{\mathbf{1}}$ in terms of the sigma
variables. The double null slicing conditions (see paper II equation (2.23))
then give
\numparts \begin{eqnarray}
    \epsilon_{ij}\Si \21i\Si \31j&=0, \label{eq:c7}\\
    \epsilon_{ij}\Si\20i\Si\30j&=0. \label{eq:c8}
\end{eqnarray} \endnumparts

We may now obtain a simplified version of these constraints by noting that
\eref{eq:c1},~\eref{eq:c3},~\eref{eq:c4},~\eref{eq:c7}
and~\eref{eq:c8} can be combined with the requirement that the
volume form be non-vanishing
\begin{equation*}
    \Si\2ai \Si\3bj\epsilon_{ab}\epsilon_{ij}\ne0,
\end{equation*}
where they reduce to the requirement that either
\begin{equation}
    \Si\202 = \Si\203 = \Si\312 = \Si\313 = 0,
\end{equation}
or
\begin{equation}
    \Si\302 = \Si\303 = \Si\213 = \Si\212 =0.
\end{equation}
These two conditions are interchanged when one swaps the $x^0$ and $x^1$
coordinates so there is no loss of generality in choosing 
the former condition which coincides with the choice of slicing
condition made in paper II. We therefore have the following constraints.
\begin{eqnarray}
    C_1\equiv \epsilon_{\alpha\beta\gamma\delta}
              \Si \1\alpha\beta \Si \2\gamma\delta=0,\\ 
    C_2\equiv \epsilon_{\alpha\beta\gamma\delta}
              \Si \1\alpha\beta \Si \3\gamma\delta=0,\\
    C_3\equiv \Si\201=0,\\
    C_4\equiv \Si\301=0,\\
    C_5\equiv \Si\202=0,\\
    C_6\equiv \Si\203=0,\\
    C_7\equiv (\Si\312)^2=0,\\
    C_8\equiv (\Si\313)^2=0.
\end{eqnarray}
Note that we have chosen to square the final two constraints as these
correspond to the existence of cyclic variables and as observed by
Goldberg {\it et al} \cite{jg&dr:92} squaring such constraints
considerably simplifies the canonical analysis.
The final constraint comes from expressing the constraint for the
conformal factor $\hat{C}$ (see paper II equation (3.10)) in terms of the
sigma variables. This results in the constraint
\begin{equation}
    C_9\equiv\epsilon_{ab}\Si \1a2\Si \1b3
        +\epsilon_{ab}\epsilon_{ij}\Si \2ai\Si \3bj -\Si\123\Si\101.
\end{equation}

We may now consider our Lagrangian to be given by~\eref{eq:Ld}, where
the variables are the components of $\Sigma_\A$ and the connection
$\Gamma^\A$, and use Lagrange multipliers to introduce the primary
constraints given above. This results in a Lagrangian density given by
\begin{equation}
\fl    \m{L}=\Si \A0i A^\A_{i,0}+\Si \A01B^\A_{1,0}
        +B^\A_0\left(D_1\Si \A01+D_i\Si \A0i\right)
        +R^\A_{\pa 1i}\Si \A1i +R^\A_{\pa 23}\Si \A23
        -\lambda^\alpha C_\alpha, \label{eq:Ld2}
\end{equation}
where $\alpha$ sums from $1,...,9$. It may be
verified that making a variation of this Lagrangian with respect to
the components of $\Sigma_A$ leads to the Einstein equations while
variation with respect to the connection leads to the structure
equations. We will not show this explicitly here as the calculations
are very similar to the derivation of these equations in the
Hamiltonian description which we give below.

\section{Hamiltonian description}
The Lagrangian density is of the form $\m
L=p^\lambda\dot{q}_\lambda-\m H$, and therefore we can see directly
from \eref{eq:Ld2} that the canonical variables are $A^\A_i$ and
$B^\A_1$, and have the respective momenta $\Si \A0i$ and $\Si
\A01$. We can also simply read off the Hamiltonian density which is given by 
\begin{equation}
 \fl   \m H=-B^\A_{0}\left(D_1\Si \A01 +D_i \Si \A0i\right)
        -R^\A_{\pa 23}\Si \A23 -R^\A_{\pa 1i}\Si \A1i
        +\lambda^\alpha C_\alpha +\xi_\A^{\pa 23}
        P^\A_{\pa 23}+\xi_\A^{\pa 1i}P^\A_{\pa 1i}+\xi^\A P_\A,
\end{equation}
where we have introduced the momenta $P^\A_{\pa 23}, P^\A_{\pa
1i},\tP_\A$ for the cyclic variables $\Si \A23, \Si \A1i,
B^\A_0$. This results in additional primary constraints which have
been introduced into the Hamiltonian using the Lagrangian multipliers
$\xi_\A^{\pa 23}, \xi_\A^{\pa 1i}, \xi_\A$. Note, that unlike paper II
we will not use the `shortcut' method of treating the cyclic
variables as if they were multipliers and eliminate them from the
canonical analysis, but will include all the variables in the
constraint analysis. This is because in the present case it is not at
all obvious that the corresponding constraints are automatically
propagated.

The canonical Poisson brackets are given by
\numparts\begin{eqnarray}
    \left\{A^\A_i(x),\Si \B0j(\ty)\right\}
        &=\delta^\A_\B\delta^j_i\,\delta(x,\ty),\\
    \left\{B^\A_1(x),\Si \B01(\ty)\right\}
        &=\delta^\A_\B\delta(x,\ty),\\
    \left\{B^\A_0(x),\tP_\B(\ty)\right\}
        &=\delta^\A_\B\delta(x,\ty),\\
    \left\{\Si \A23(\tx),P^\B_{\pa 23}(y)\right\}
        &=\delta^\A_\B\delta(\tx,y),\\
    \left\{\Si \A1i(\tx),P^\B_{\pa 1j}(y)\right\}
        &=\delta^\A_\B\delta^i_j\,\delta(\tx,y).
\end{eqnarray}\endnumparts
We have a total of twenty one primary constraints introduced into the
Hamiltonian. Following the Dirac-Bergmann algorithm we propagate the
primary constraints. The equations for $\dot{C}_1$, $\dot{C}_2$ and 
$\dot{C}_9$ are rather complicated so we do not give them explicitly
however they define the multipliers $\xi_\2^{\pa 23}$, $\xi_\3^{\pa 23}$ and
$\xi_\1^{\pa 23}$ respectively. The remaining constraints give 
\numparts\begin{eqnarray}
\fl    \dot{C}_3&= \left(D_i \Si\A1i
        +2\eta^c_{\pa \A\B}B^\B_0 \Si\C01\right)\delta^\A_\2,\\
\fl    \dot{C}_4&=\left(D_i\Si \A1i
        +2\eta^c_{\pa \A\B}B^\B_0\Si \C01\right)\delta^\A_3,\\
\fl    \dot{C}_{5,6}&\equiv\dot{\tilde{\Sigma}}_\2^{\pa 0i}
        =-2\eta^\C_{\pa \2\A}B^\B_0\Si \C0i
        -\tilde{\Sigma}_{\2\pa,1}^{\pa 1i}
        +2\eta^\A_{\pa \B\2}B^\B_1\Sigma_\A^{\pa 1i} +D_j\Si \2ij,\\
\fl    \dot{C}_{7,8}&\equiv\dot{\Sigma}_\3^{\pa 1i}=\xi_\3^{\pa 1i}.
\end{eqnarray}\endnumparts
From which we see that the equations $\dot{C}_{7}$ and $\dot{C}_8$ define
the multipliers $\xi_\3^{\pa 1i}$. 
We also need to propagate the momenta conjugate to the cyclic
variables. This gives
\numparts\begin{eqnarray}
\fl    \dot{\tP}_\A&=D_1\Si \A01 +D_i \Si \A0i,\label{sig:dPA}\\
\fl    \dot{P}^\A_{\pa 23}&=R^\A_{\pa 23}
        -\delta^\A_1(\Si \202\lambda^1+\Si \301\lambda^2
        -\Si\101\lambda^9)-\delta^\A_2\Si\101\lambda^1
        -\delta^\A_3\Si \101\lambda^2,\label{sig:dPA23}\\
\fl    \dot{P}^\A_{\pa 12}&=R^\A_{\pa 12}
        -\delta^\A_\1(\Si \203\lambda^1+\Si \303\lambda^2
            -\Si\103\lambda^9)
        -\delta^\A_\2(\Si\103\lambda^1-\Si\303\lambda^9)\nonumber\\
\fl        &-\delta^\A_\3(\Si \103\lambda^2-\Si\203\lambda^9
            +2\Si\312\lambda^7),\label{sig:dPA12}\\
\fl    \dot{P}^\A_{\pa 13}&=R^\A_{\pa 13}
        +\delta^\A_\1 (\Si \202\lambda^1+\Si \302\lambda^2
            -\Si\102\lambda^9)
        +\delta^\A_\2(\Si\102\lambda^1-\Si\302\lambda^9)\nonumber\\
\fl        &+\delta^\A_\3(\Si\102\lambda^2-\Si\202\lambda^9
            +2\Si\313\lambda^8).\label{sig:dPA13}
\end{eqnarray}\endnumparts
Equations \eref{sig:dPA23}--\eref{sig:dPA13} define the multipliers $\lambda^1$, $\lambda^2$
and $\lambda^9$ through the equations
$\lambda^1\Si\101=R^\2_{\pa 23}$, $\lambda^2\Si\101=R^\3_{\pa 23}$ and
$\lambda^9\Si\101\approx -R^\1_{\pa 23}$. This leaves thirteen
secondary constraints given by
\numparts\begin{eqnarray}
    \dot{C}_3&=\left(D_i\Si \A1i+2\eta^\C_{\pa \A\B}B^\B_0\Si
        \C01\right)\delta^\A_\2=0,\label{sig:C3dot}\\
    \dot{C}_4&=\left(D_i\Si \A1i+2\eta^\C_{\pa \A\B}B^\B_0\Si
        \C01\right)\delta^\A_\3=0,\label{sig:C4dot}\\
    \dot{\tilde{\Sigma}}_\2^{\pa 0i}
        &=-2\eta^\C_{\pa \2\A}B^\B_0\Si \C0i
        -\Sigma_{\2\pa,1}^{\pa 1i}
        +2\eta^\A_{\pa\B\2}B^\B_1\Sigma_\A^{\pa 1i}
        +D_j\Si \2ij=0,\label{sig:C7&8dot} \\
\dot{P}^\1_{\pa 1i}&\approx R^\1_{\pa 1i}\Si \101
        -R^\1_{\pa ij}\Si \10j
        -R^\2_{\pa ij}\Si \20j
        -R^\3_{\pa ij}\Si \30j\approx0,\label{sig:P11jdot}\\
    \dot{P}^\2_{\pa 1i}&\approx R^\2_{\pa 1i}\Si \101
        -R^\2_{\pa ij}\Si \10j
        -R^\1_{\pa ij}\Si \30j=0,\label{sig:P21jdot}\\
    \dot{P}^\3_{\pa 1i}&\approx R^\3_{\pa 1i}\Si \101
        -R^\3_{\pa ij}\Si \10j
        -R^\1_{\pa ij}\Si \20j=0,\label{sig:P31jdot}\\
    \dot{\tP}_\A&=D_1\Si \A01 +D_i \Si \A0i=0.\label{sig:PAdot}
    \end{eqnarray}\label{sig:secondary}\endnumparts

The Dirac-Bergmann algorithm then requires us to propagate these
secondary constraints to see if any further tertiary  constraints
arise. Before this is done we rewrite the secondary constraint
\eref{sig:P31jdot} by looking at its components. This gives
\numparts\begin{eqnarray}
    \Si\30i\dot{P}^\3_{\pa 1i}&=\Si\30i(R^\3_{\pa 1i}\Si \101
        -R^\3_{\pa ij}\Si\10j-R^\1_{\pa ij}\Si\20j),\label{sig:sa}\\
    \Si\11i\dot{P}^\3_{\pa 1i}&=\Si\11i(R^\3_{\pa 1i}\Si \101
        -R^\3_{\pa ij}\Si\10j-R^\1_{\pa ij}\Si\20j)\label{sig:sb}.
\end{eqnarray}\endnumparts
As we show in section 4 the five equations \eref{sig:P11jdot},
\eref{sig:P21jdot} and \eref{sig:sa} define five of the Einstein
equations and are therefore conserved by the Bianchi
identity. Propagation of \eref{sig:PAdot} is identically zero for
$A=0$, and defines the multipliers $\lambda^3$ and $\lambda^4$ for the
remaining values of $A$. Equations \eref{sig:C3dot} and
\eref{sig:C4dot} define the multipliers $\xi^\3$ and $\xi^\2$ when
propagated. Propagation of \eref{sig:C7&8dot} define the multipliers
$\xi_\2^{\pa 1i}$, and finally propagation of \eref{sig:sb} leads to
a multiplier equation giving the multipliers $\xi_\1^{\pa
1i}$. Therefore no additional constraints arise through the propagation of
the secondary constraints.

Having dealt with the constraints we next calculate the equations of
motion. Evolving $A^\A_i$ gives 
\begin{equation}
\begin{array}{ll}
    \dot{A}^\A_i&=D_iB^\A_0-\delta^\A_\1\epsilon_{ij}
        (\Si \20j\lambda^1+\Si \30j\lambda^2-\Si\11j\lambda^9) \\
        &-\delta^\A_\2\epsilon_{ij}(\Si\11j\lambda^1
            -\Si\31j\lambda^9)\\
        &-\delta^\A_\3\epsilon_{ij}(\Si\11j\lambda^2
            -\Si \21j\lambda^9),
\end{array}
\end{equation}
which results in
\numparts\label{sig:eofm}\begin{eqnarray}
    & \dot{A}^\1_i=D_iB^\1_0
      -\left(R^\2_{\pa ij}\Si \20j+R^\3_{\pa ij}\Si \30j
      +R^\1_{\pa ij}\Si \11j\right)/\Si \101,\\
    &\dot{A}^\2_i=D_iB^\2_0-(R^\2_{\pa ij}\Si \11j
        +R^\1_{\pa ij}\Si\31j)/\Si \101,\\
    &\dot{A}^\3_i=D_iB^\3_0-
        (R^\3_{\pa ij}\Si \11j+R^\1_{\pa ij}\Si\21j)/\Si \101.
\end{eqnarray}
The remaining equations of motion are given by
\begin{eqnarray}
    \dot{\tilde{\Sigma}}^{\pa 01}_\A
        =2\eta^\C_{\pa \B\A}B^\B_0\Si \C01+D_i\Si \A1i,\\
    \dot{\tilde{\Sigma}}_\A^{\pa 0i}=
        2\eta^\C_{\pa \B\A}B^\B_0\Si \C0i
        -D_1\Si \A1i-D_j\Si \A ji,\\
    \dot{B}^\A_1=D_1B^\A_0 +\delta^\A_\1
        (\Si \223\lambda^1+\Si \323\lambda^2-\Si\123\lambda^9)\\
        +\delta^\A_\2(\Si\1ij\lambda^1+\lambda^3)
        +\delta^\A_\3(\Si\1ij\lambda^2+\lambda^4).
\end{eqnarray}\endnumparts

\section{Einstein Equations}

Since we have now obtained all the constraints and equations of motion we
are now in a position to derive the Einstein equations. We use five of the
secondary equations (\ref{sig:P11jdot}, \ref{sig:P21jdot} and
\ref{sig:sa}), together with the expression for the Einstein tensor in
terms of the self-dual curvature ( see equation (5.1) of paper II) to
obtain the following five Einstein equations.  
\numparts
\begin{eqnarray}
\dot{P}^\2_{\pa 1j}\Si \21j= \left(R^\2_{\pa 1j}\Si \101 +R^\2_{\pa ij}\Si
\10i+R^\1_{\pa ij}\Si\30i\right) \Si \21j\approx0 
\Longleftrightarrow G^\0_{\pa \0}&\approx 0,\\ 
\dot{P}^\3_{\pa 1j}\Si \30j= \left(R^\3_{\pa 1j}\Si \101 
+R^\3_{\pa ij}\Si \10i+R^\1_{\pa ij}\Si\20i\right) \Si \30j\approx0 
\Longleftrightarrow G^\0_{\pa \1}&\approx 0,\\ 
\dot{P}^\1_{\pa 1j}\Si \21j=\left(R^\1_{\pa 1j}\Si \101 
+R^\1_{\pa ij}\Si\10i+R^\3_{\pa ij}\Si \30i\right) \Si \21j\approx0 
\Longleftrightarrow G^\0_{\pa \2}&\approx 0,\\ 
\dot{P}^\1_{\pa 1j}\Si \30j\approx \left(R^\1_{\pa 1j}\Si \101 
+R^\1_{\pa ij}\Si\10i +R^\3_{\pa ij}\Si \30i\right) \Si\30j\approx0
\Longleftrightarrow G^\0_{\pa \3}&\approx 0,\\
\dot{P}^\2_{\pa 1j}\Si\30j\approx \left(R^\2_{\pa 1j}\Si\101
+ R^\2_{\pa ij}\Si \10i +R^\1_{\pa ij}\Si\30i\right) \Si\30j\approx0
\Longleftrightarrow G^\2_{\pa \3}&\approx 0.
\end{eqnarray}
\endnumparts
The remaining Einstein equations are obtained from the equations of
motion from which we obtain
\numparts\begin{eqnarray}
\fl    \dot{A}^\2_i\Si \21i\approx -\left(R^\2_{\pa 0i}\Si \101
        +R^\1_{\pa ij}\Si\31j+R^\2_{\pa ij}\Si \11j\right)
        \Si \21i\approx0\Longleftrightarrow G^\1_{\pa \0}\approx0,\\
\fl    \dot{A}^\1_i\Si \21i\approx -\left(R^\1_{\pa 0i}\Si \101
        +R^\1_{\pa ij}\Si\11j+R^\2_{\pa ij}\Si \21j\right)
        \Si\21i\approx0 \Longleftrightarrow G^\1_{\pa\2}\approx 0,\\
\fl    \dot{A}^\1_i\Si \30i\approx -\left(R^\1_{\pa 0i}\Si \101
        +R^\1_{\pa ij}\Si\11j+R^\2_{\pa ij}\Si \21j\right)
        \Si\30i\approx0 \Longleftrightarrow G^\1_{\pa \3}\approx0,\\
\fl    \dot{A}^\3_i\Si \21i\approx -\left(R^\3_{\pa 0i}\Si \101
        +R^\1_{\pa ij}\Si\21j+R^\3_{\pa ij}\Si \10j\right)
        \Si\21i\approx0 \Longleftrightarrow G^\3_{\pa \2}\approx0,\\
\fl    \dot{A}^\1_i\Si \10i+\dot{B}^\1_1\Si \101\approx
        -\left(R^\1_{\pa 0i}\Si \101 +R^\1_{\pa ij}\Si\11j
        +R^\2_{\pa ij}\Si \21j\right)\Si \10i \nonumber \\
\fl    -\left(R^\1_{\pa 01}\Si\101 -R^\1_{\pa 23}\Si\123
        -R^\2_{\pa23}\Si\223 -R^\3_{\pa23}\Si\323\right)\Si\101
        \approx0 \Longleftrightarrow G^\3_{\pa \3}\approx 0.
\end{eqnarray}\endnumparts
We have therefore shown that we can derive all the Einstein equations from
the constraint and evolution equations. The structure equations are
obtained in exactly the same way as in paper II so we do not repeat the
calculation here. The final stage is to ascertain which constraints are
first class and then calculate the algebra of first class constraints.

\section{First class constraints}
We now move on to the next stage of the Dirac-Bergmann algorithm and
calculate the first class constraints. This is harder than in paper II
because we do not have available such an obvious geometric
interpretation of the constraints.  Fortunately some of the secondary
constraints are the same as the corresponding
secondary equations obtained in paper I and we can therefore adapt
them by adding appropriate combinations of the other constraints in
the same way as we did in paper II to obtain four first class
constraints.  However because we are not using the `shortcut' method
but including the cyclic variables in our analysis we have to add
extra terms to the three constraints $\psi_1$, and  $\psi_p$ to
ensure they are first class. This results in the following constraints.
\numparts\begin{eqnarray}
    \psi_1:= B^\A_{1,1}\Si \A01 +A^\A_{i,1}\Si \A0i
        -(B^\A_1\Si \A01)_{,1}-(B^\A_1\Si \A0i)_{,i} \nonumber\\
        +B^\A_{0,1}\tP_\A -\Si \A23 P^\A_{\pa 23,1}
        -\Si \A1i P^\A_{\pa 1i,1}
        +(\Si \A1i P^\A_{\pa 1i})_{,1}\approx0,\label{sig:fpsi1}\\
    \psi_p:= B^\A_{1,j}\Si \A01 +A^\A_{i,j}\Si \A0i
        -(A^\A_j \Si \A0i)_{,i} -(A^\A_j\Si \A01)_{,1} \nonumber\\
        +B^\A_{o,j}\tP_\A
        -\Si \A23 P^\A_{\pa 23,j} -\Si \A1i P^\A_{\pa 1i,j}
        +(\Si \A1i P^\A_{\pa 1j})_{,i}\approx0.
        \label{sig:fpsii}
\end{eqnarray}\endnumparts
We also need to modify the Gauss constraint to take account of the
extra cyclic variables. This results in
\begin{equation}
\fl    \m G_\1:=D_1\Si \101+D_i\Si \10i
        +2\eta^\C_{\pa \B\1}P^\B_{\pa 1i}\Si \C1i
        +2\eta^\C_{\pa \B\1}P^\B_{\pa 23}\Si \C23
        -2\eta^\C_{\pa\B\1}B^\B_0\tP_\C \approx0.\label{sig:fG}
\end{equation}
Because we are including the cyclic variables in the analysis it turns out
that we also obtain an additional two first class constraints which are 
given by
\numparts\begin{eqnarray} 
P^\1_{\pa 23}&=0,\label{sig:fP123}\\
\tP_\1&=0.\label{sig:fP1}
\end{eqnarray}\endnumparts
We can show that there are no further first class constraints so that we
have a total of six first class given by \eref{sig:fpsii}--\eref{sig:fP1}.
The remaining constraints are second class which for completeness we list 
below;
\numparts
\begin{eqnarray} 
C_\alpha=0, \quad\alpha=1,...,9  \label{secondf}\\
P^\A_{\pa 1i}=0, \\
P^\2_{\pa 23}=0,  \\
P^\3_{\pa 23}=0, \\
\tP_\2=0, \\
\tP_\3=0, \\ 
\left(D_1\Si \A01 +D_i \Si \A0i\right)\delta^\A_\2=0, \\
\left(D_1\Si \A01 +D_i \Si \A0i\right)\delta^\A_\3=0,\\
\left(D_i\Si \A1i+2\eta^\C_{\pa \A\B}B^\B_0\Si \C01\right)\delta^\A_\2=0, \\
\left(D_i\Si A1i+2\eta^\C_{\pa \A\B}B^\B_0\Si \C01\right)\delta^\A_\3=0,\\
\left(R^\A_{\pa 1j}\Si \101+R^\A_{\pa ij}\Si \10i\right)\delta^\A_\2=0, \\
\left(R^\A_{\pa 1j}\Si \101+R^\A_{\pa ij}\Si \10i\right)\delta^\A_\3=0,\\
D_1\Si \2i1 +D_j\Si \2ij-2\eta^\C_{\pa \2\A}B^\B_0\Si \C0i=0, \\
\Si\11i(R^\3_{\pa 1i}\Si \101
-R^\3_{\pa ij}\Si\10j-R^\1_{\pa ij}\Si\20j)=0. \label{secondl}
\end{eqnarray}
\endnumparts 
Because we are not using the `shortcut' method but working
with a phase space which includes all the cyclic variables we have a rather
large phase space consisting of 42 functions. These functions are subject
to 6 first class and 28 second class constraints leaving 2 dynamical
degrees of freedom per hypersurface point as is appropriate on a null
surface \cite{sachs}, \cite{goldberg}, \cite{goldsot}, \cite{rp:80}.

Having obtained all the first class constraints we calculate the first
class algebra by computing the Poisson brackets of the constraints
with each other. To do this we first obtain smooth versions of the
constraints by smearing them with smooth functions.
We start by considering $\psi_i$. Let $\tilde F$ be a vector field with
components tangent to $\{S\}$ and define a smeared version of the $\psi_i$
constraint by
\begin{equation}
\tilde\Psi(\tilde F)=\int \tilde F^i\psi_i \ud^3x.
\end{equation}
We next consider $\psi_1$. This time we let $\hat F$ be a vector field with
components tangent to the null generators of $\Sigma_\0$ and defined a
smeared version of the $\psi_1$ constraint by
\begin{equation}
\hat\Psi(\hat F)=\int \hat F^1\psi_1 \ud^3x.
\end{equation}
In a similar way we define a smeared versions of the remaining constraints by
\begin{eqnarray}
G(f)=\int f \m G_1 \ud^3x, \\
P(f)=\int f \tilde P_\1 \ud^3x, \\
\Pi(\tilde f)=\int \tilde f P^1_{\pa 23} \ud^3x
\end{eqnarray}
where $f$ is a scalar field and $\tilde f$ a scalar density. 
We now obtain the first class algebra by calculating the Poisson brackets
of all the first class constraints with each other. Below we show only
those terms that are not strongly zero
\numparts
\begin{eqnarray}
\left\{\tilde\Psi(\tilde R),\tilde\Psi(\tilde S)\right\}=
\tilde\Psi(\m L_{\tilde R}\tilde S),\\
\left\{\tilde\Psi(\tilde R),\hat\Psi(\hat S)\right\}=
\hat\Psi(\m L_{\tilde R}\hat S),\\
\left\{\hat\Psi(\hat R),\hat\Psi(\hat S)\right\}=
\hat\Psi(\m L_{\hat R}\hat S),\\
\left\{\tilde\Psi(\tilde R),G(s)\right\}=
G(\m L_{\tilde R}s),\\
\left\{\hat\Psi(\hat R),G(s)\right\}=
G(\m L_{\hat R}s),\\
\left\{\tilde\Psi(\tilde R),\Pi(\tilde s)\right\}=
\Pi(\m L_{\tilde R}\tilde s),\\
\left\{\hat\Psi(\tilde R),\Pi(\tilde s)\right\}=
\Pi(\m L_{\tilde R}\tilde s),\\
\left\{\tilde\Psi(\tilde R),P(\tilde s)\right\}=
P(\m L_{\tilde R}\tilde s),\\
\left\{\hat\Psi(\tilde R),P(\tilde s)\right\}=
P(\m L_{\tilde R}\tilde s).
\end{eqnarray}
\endnumparts
As in paper II we have chosen to keep $\psi_1$ and $\psi_i$ separate to
illustrate the 2+2 structure of the constraint algebra. However they may be
combined to give $\psi_A$, where $(\psi_A)=(\psi_1, \psi_2, \psi_3)$. This
may be smeared with a general vector field $F$ on $\Sigma_0$ to give
\begin{equation}
\Psi(F)=\int F^A\psi_A \ud^3x.
\end{equation}
The constraint algebra then has the more compact form
\numparts
\begin{eqnarray}
\left\{\Psi(R),\Psi(S)\right\}=
\Psi(\m L_{R} S),\\
\left\{\Psi(R),G(s)\right\}=
G(\m L_{R}s),\\
\left\{\Psi(R),\Pi(\tilde s)\right\}=
\Pi(\m L_{R} \tilde s),\\
\left\{\Psi(R),P(s)\right\}=
P(\m L_{R} s).
\end{eqnarray}
\endnumparts

Having calculated the first class algebra, we give a geometrical
interpretation of the first class constraints by calculating the
infinitesimal transformations the constraints generate.  However we
will not consider the constraints $\tP_1=0$ and $P^1_{\pa 23}=0$ (which
come from the cyclic variables) as they just indicate the gauge freedom
to choose the variables $B^1_0$ and $\Si 123$.  We start with the
constraint $\psi_i$. Calculating the Poisson brackets of this with the
other variables gives
\numparts\begin{eqnarray}
    \delta A^\A_{\pa i}&=\left\{A^\A_i,\tilde\Psi(G)\right\}=
        \m L_{G} A^\A_{\pa i},\\
    \delta B^\A_1&=\left\{B^\A_0,\tilde\Psi(G)\right\}=\m L_{G} B^\A_{\pa 1},\\
    \delta B^\A_0&=\left\{B^\A_0,\tilde\Psi(G)\right\}=\m L_{G} B^\A_{\pa 0},\\
    \delta \Si \A23&=\left\{\Si \A23,\tilde\Psi(G)\right\}=\m L_{G} \Si \A23,\\
    \delta \Si \A1i&=\left\{\Si \A1i,\tilde\Psi(G)\right\}=\m L_{G} \Si \A1i.
\end{eqnarray}\label{ash:infinit_i}\endnumparts
From this and we see that, as in paper II,  $\psi_i$ generates 
diffeomorphisms in the two surface $\{S\}$.

We next consider the constraint $\psi_1$.
\numparts\begin{eqnarray}
    \delta A^\A_{\pa i}&=\left\{A^\A_i,\hat\Psi(G)\right\}=
        \m L_{G} A^\A_{\pa i},\\
    \delta B^\A_1&=\left\{B^\A_0,\hat\Psi(G)\right\}=\m L_{G} B^\A_{\pa 1},\\
    \delta B^\A_0&=\left\{B^\A_0,\hat\Psi(G)\right\}=\m L_{G} B^\A_{\pa 0},\\
    \delta \Si \A23&=\left\{\Si \A23,\hat\Psi(G)\right\}=\m L_{G} \Si \A23,\\
    \delta \Si \A1i&=\left\{\Si \A1i,\hat\Psi(G)\right\}=\m L_{G} \Si \A1i.
\end{eqnarray}\label{ash:infinit_1}\endnumparts
From this we see that this constraint generates the diffeomorphism
along the null generators on $\Sigma_\0$.

Finally we look at the commutators with the modified Gauss constraint
$\m G_\1$. These give
\numparts\begin{eqnarray}
    \delta A^\A_i&=\left\{A^\A_i,G(g)\right\}=-g_{,i}\delta^\A_\1
        -2gA^\2_i\delta^\A_\2 +2gA^\3_i\delta^\A_\3,\\
    \delta B^\A_1&=\left\{B^\A_1,G(g)\right\}=-g_{,1}\delta^\A_\1
        -2gB^\2_1\delta^\A_\2 +2gB^\3_1\delta^\A_\3,\\
    \delta B^\A_0&=\left\{B^\A_0,G(g)\right\}=
        -2gB^\2_0\delta^\A_\2 +2gB^\3_0\delta^\A_\3,\\
    \delta \Si \A23&=\left\{\Si \A23,G(g)\right\}=
        -2g\Si \223\delta^\A_\2 +2g\Si \323\delta^\A_\3,\\
    \delta \Si \A1i&=\left\{\Si \A1i,G(g)\right\}=
        -2g\Si \21i\delta^\A_\2 +2g\Si \31i\delta^\A_\3.
\end{eqnarray} \endnumparts
Comparing these infinitesimal transformations with the effect of the
self-dual spin and boost transformations (see equation 
(7.15) of paper II) we see that $\m G_\1$ again generates the self-dual spin 
and boost transformations for these variables.

\section{Conclusion}
In this paper we have obtained a 2+2 double null Hamiltonian description of
General Relativity using only the complex $SO(3)$ connection and the
components of the complex densitised self-dual bivectors $\Sigma_\A$. We have
carried out the general canonical analysis of this system and obtained the
first class constraint algebra entirely in terms of the self-dual
variables. All the first class constraints have a simple geometrical
interpretation. The constraint $\psi_i$ generates the diffeomorphisms in
the two surface $\{S\}$, and the constraint $\psi_1$ generates the
diffeomorphisms along the null generators of $\Sigma_\0$, so that these
constraints can be combined to give $\psi_A$ which is just the momentum
constraint in this description. The modified Gauss constraint $\m G_\1$
generates the self-dual spin and boost transformations, while the final two
constraints correspond to cyclic variables and express the freedom to
freely choose the corresponding canonical variables. The choice of
variables we have made makes it possible to compare the results of using a
double null evolution with both the standard 3+1 approach using Ashtekar
variables \cite{aa:87} and the null approach of Goldberg {\it et al}
\cite{jg&dr:92}. In particular one can see how the form of $\psi_A$ is very
similar to that of the adapted first class version of the momentum
constraint
\begin{equation}
\tilde\Pi^\nu_{\i\bj}\m R_{\mu\nu}^{\pa\pa \i\bj}-
\m A^{\i\bj}_{\nu}D_\mu\left(\tilde \Pi^\mu_{\i\bj}\right)=0
\end{equation}
which is used in the standard Ashtekar approach.

The next step of the canonical quantisation process is to eliminate the
second class constraints by replacing the Poisson brackets by Dirac
brackets \cite{Dirac}.These are modified versions of the Poisson brackets
such that the Dirac bracket between any of the second class constraints and
any other variable vanishes identically. The details of the analysis of the
second class constraints and the construction of the Dirac brackets will be
presented elsewhere, but we outline the general procedure below. We start
by introducing a vector $K$ whose components $K_I$, $I=1,\ldots 28$ are the
second class constraints given by equation \eref{secondf}--\eref{secondl}.
We then calculate the Poisson bracket matrix $C$ for the second class
constraints whose components are given by
\begin{equation}
C_{IJ}=\{C_I,C_J\}\ . \label{secondalg}
\end{equation}
Although in principle $C$ is a large matrix, the form of the second class
constraints mean that, as in paper II, many of the entries are identically
zero. 

Having calculated $C$ the Dirac brackets are then given by
\begin{equation}
\{F,G\}_D=\{F,G\}-\sum_{J,K}\{F,C_J\}C^{-1 JK}\{C_K,G\},
\end{equation}
where $C^{-1}$ is the inverse of the matrix given by \eref{secondalg}.
Note that the invertibility of $C$ is guaranteed by the independence of the
second class constraints. Rather than calculating the Dirac brackets an
alternative procedure would be to impose a suitable gauge condition and use
this to explicitly solve for the second class constraints and hence obtain 
a Hamiltonian which only depended on the true dynamical degrees of
freedom (see e.g. Goroff and Schwartz \cite{goroffschwartz})

As well as working with entirely self-dual variables another important
feature of the approach taken in this paper is the use of a double null
evolution. This has the advantage over choosing a null hypersurface that
the projection operation is well defined. It also has the advantage over
the standard 3+1 approach that the Hamiltonian
constraint is second class rather than first class. This has the important
effect that the first class constraint algebra also forms a Lie
algebra. As described above the next step of the standard canonical
quantisation process would be to calculate the Dirac brackets. However an
alternative approach would be to look at the corresponding loop variables
on the null hypersurface. This might provide a link between the Ashtekar
approach to loop quantum gravity (see for example \cite{ashlewandowski})
and the work of Iyer, Kozameh and Newman on formulations of Einstein's
equations based on the holonomy of null surfaces \cite{IKN}.

\end{document}